\documentclass[fleqn,10pt]{wlscirep}
\usepackage{epstopdf}
\usepackage{textcomp}
\usepackage{gensymb}
\usepackage[utf8]{inputenc}
\usepackage{nicefrac}
\usepackage{cases}
\usepackage{subcaption}
\usepackage{cite}
\title{Universal classification of twisted, strained and sheared graphene moir$\text{\'e}$ superlattices}

\author[1,2,3,*]{Artaud A.}
\author[1,2]{Magaud L.}
\author[1,3]{Le Quang T.}
\author[1,2]{Guisset V.}
\author[1,2]{David P.}
\author[1,3]{Chapelier C.}
\author[1,2]{Coraux J.}
\affil[1]{Univ. Grenoble Alpes, F-38000 Grenoble, France}
\affil[2]{CNRS, Inst NEEL, F-38000 Grenoble, France}
\affil[3]{CEA, INAC-PHELIQS, F-38000 Grenoble, France}

\affil[*]{alexandre.artaud@cea.fr}

\keywords{graphene, moir$\text{\'e}$, STM, DFT}

\begin{abstract}
Moir$\text{\'e}$ superlattices in graphene supported on various substrates have opened a new avenue to engineer graphene's electronic properties. Yet, the exact crystallographic structure on which their band structure depends remains highly debated. In this scanning tunneling microscopy and density functional theory study, we have analysed graphene samples grown on multilayer graphene prepared onto SiC and on the close-packed surfaces of Re and Ir with ultra-high precision. We resolve small-angle twists and shears in graphene, and identify large unit cells comprising more than 1,000 carbon atoms and exhibiting non-trivial nanopatterns for moir$\text{\'e}$ superlattices, which are commensurate to the graphene lattice. Finally, a general formalism applicable to any hexagonal moir$\text{\'e}$ is presented to classify all reported structures.
\end{abstract}
\begin{document}

\flushbottom
\maketitle
\thispagestyle{empty}

\section*{Introduction}

Graphene is a two-dimensional crystal with honeycomb structure, whose peculiar electronic properties have raised considerable interest in the past few years. Indeed, its electronic bands cross at the $K$ and $K'$ corners of the Brillouin zone, giving rise to a linear energy dispersion of its quasiparticles close to the Fermi level\cite{wallace_band_1947}{}. Moreover, the bipartite nature of graphene's lattice, with two triangular carbon sub-lattices (A and B), confers unique properties to these quasiparticles. By analogy to quantum electrodynamics\cite{semenoff_condensed-matter_1984}{}, a sublattice-related quantum number, so-called pseudo-spin, equivalent to the spin of Dirac fermions is defined\cite{novoselov_two-dimensional_2005}{}. For these reasons, the conical electronic bands around the $K$ and $K'$ points of the Brillouin zone are called Dirac cones.

Such exotic electronic properties are predicted for pristine graphene, but are altered when graphene is supported by a substrate. Indeed, due to the structural mismatch between graphene and its support, graphene has periodically varying stacking configurations with its substrate. This effect modulates the graphene-substrate interaction and distance\cite{wang_chemical_2008,brako_graphene_2010,busse_graphene_2011}{}, over a so-called moir$\text{\'e}$ periodicity, which can range from $\sim1$ to $\sim15~\mathrm{nm}$. Depending on the interaction between graphene and the substrate, the moir$\text{\'e}$ has a dramatic impact on graphene's properties. Some substrates show only a weak interaction dominated by van der Waals forces, such as graphene on hexagonal boron nitride\cite{dean_boron_2010} or multilayer graphene on the carbon face of SiC\cite{hass_structural_2007}{}. In this case, the graphene-substrate distance is of the order of $3.4~\text{\AA}$ (Refs.~\citen{hass_structural_2007,haigh_cross-sectional_2012}), very close to the value $3.3539~\text{\AA}$ of highly oriented pyrolytic graphite (HOPG)\cite{lide_crc_1994}{}, and graphene's electronic properties are mostly preserved\cite{orlita_approaching_2008,haigh_cross-sectional_2012}{}. In these systems, the moir$\text{\'e}$ acts as a smooth superpotential that varies slowly compared to the one associated to carbon atoms. Still, it implies a larger unit cell, in other words replicas of the electronic bands (mini-bands), which are associated with replica Dirac cones, reduced Fermi velocity\cite{pletikosic_dirac_2009,ortix_graphene_2012,yankowitz_emergence_2012,ohta_evidence_2012}, with either superlattice Dirac cones\cite{park_new_2008,ortix_graphene_2012,yankowitz_emergence_2012,ponomarenko_cloning_2013} or mini-gaps\cite{pletikosic_dirac_2009,hunt_massive_2013} at the moir$\text{\'e}$ Brillouin zone boundary. Such preserved properties make this system an ideal playground to investigate quantum phases arising in periodic two-dimensional electron gases subjected to an external magnetic field\cite{dean_hofstadters_2013,ponomarenko_cloning_2013,hunt_massive_2013}{}. In bilayer graphene samples, Van Hove singularities and electron localization also emerge from the coexistence of the Dirac cones of each layer\cite{luican_single-layer_2011,trambly_de_laissardiere_numerical_2012}{}. Therefore, in low-interaction systems, tuning moir$\text{\'e}$ superlattices is a mean to tailor graphene's electronic properties.

Other surfaces interact more strongly with graphene, and are prone to exchange electrons with it, establishing partially covalent bonds. Graphene-substrate bonding then implies both van der Waals forces (physisorption) and partial covalent bonding (chemisorption), and is modulated along the moir$\text{\'e}$ periodicity\cite{brako_graphene_2010,busse_graphene_2011,stradi_role_2011,gao_epitaxial_2011,tonnoir_induced_2013}{}. Graphene is thus nanorippled with short graphene-substrate distances where the tendency to covalent bonding is more prominent. Nanorippling amplitudes varying from $0.03$ (on Pt(111)\cite{ugeda_point_2011}) to $1.6~\text{\AA}$ (on Re(0001)\cite{tonnoir_induced_2013}) have been reported depending on the strength of the graphene-substrate interaction\cite{tetlow_growth_2014}{}. Systems with strong nanorippling amplitudes usually have valence and conduction bands without Dirac fermion character\cite{sutter_electronic_2009,papagno_large_2012}{}. For all those metals, however, the moir$\text{\'e}$ modulation of graphene's electronic properties goes along with a modulation of its chemical reactivity, inducing preferential sites for adsorbtion or functionalization. This renders possible to use moir$\text{\'e}$ superlattices as a template for self-organized arrays of metallic clusters\cite{ndiaye_structure_2008,donner_structural_2009} or molecules\cite{zhang_assembly_2011,yang_moleculesubstrate_2012}{}. For gr/Ir, they can in turn influence the electronic properties of graphene, for instance opening a band gap\cite{balog_bandgap_2010,papagno_large_2012}{} or tuning the Fermi level and velocity\cite{rusponi_highly_2010,scardamaglia_graphene-induced_2013}{}. Understanding moir$\text{\'e}$ superlattices is then useful to shape such templates.

\begin{figure}[t]
\centering
\includegraphics{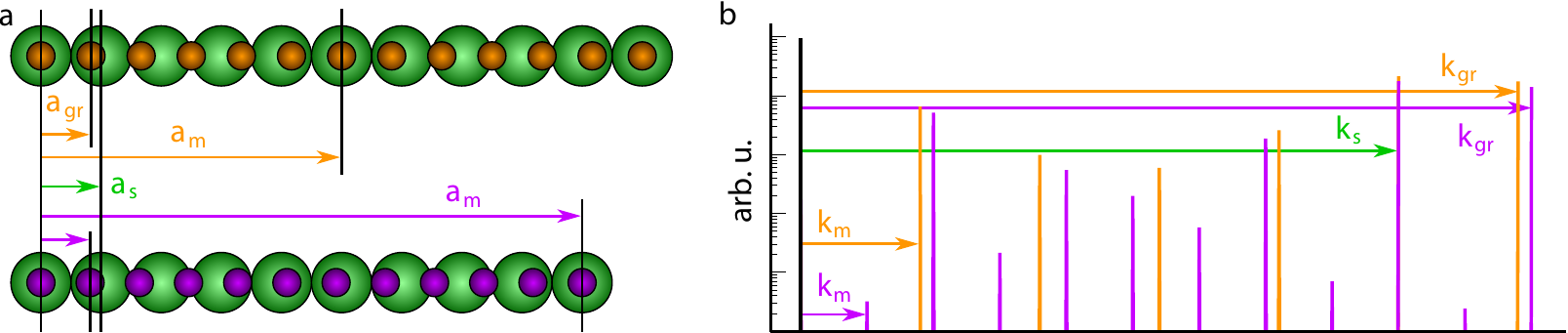}
\caption{\textbf{Moir$\text{\'e}$ superlattice and beatings}: (\textbf{a})~Ball model of a chain of (small) carbon atoms in graphene on top of a chain of (large) atoms from the support, both having different lattice parameters $a_\mathrm{gr}$ and $a_\mathrm{s}$, whose commensurability define a moir$\text{\'e}$ superlattice with period $a_\mathrm{m}$. For 6 graphene periods matching 5 support periods (orange), a single beating occurs within the moir$\text{\'e}$ period, and the fast Fourier transform (FFT) reveals a fundamental harmonic defined by $k=\nicefrac{(k_\mathrm{gr}-k_\mathrm{s})}{(6-5)}$ (\textbf{b}). For 11 graphene periods matching 9 support periods (purple), two beatings occur within the moir$\text{\'e}$ period, with similar stacking configurations at the edges and at the middle of the ball model. The corresponding FFT reveals a fundamental harmonic at $k=\nicefrac{(k_\mathrm{gr}-k_\mathrm{s})}{(11-9)}$ (\textbf{b}).}
\label{fig:moire-sketch}
\end{figure}

The usual structural model used to describe moir$\text{\'e}$s assumes (Fig.~\ref{fig:moire-sketch}a) that a single moir$\text{\'e}$ beating occurs within a moir$\text{\'e}$ period and that the graphene and moir$\text{\'e}$ lattices are commensurate (integer multiples of their lattice parameters can be found which are equal). This superlattice model was for instance often used to describe gr/Ir, but eventually proved too restrictive to describe the variety of situations observed in experiments (see Refs.~\citen{loginova_defects_2009,vo-van_epitaxial_2011,hattab_growth_2011,meng_multi-oriented_2012,jean_effect_2013,hamalainen_structure_2013} for gr/Ir) depending on growth conditions\cite{blanc_local_2012} and sample history\cite{hattab_interplay_2012,jean_effect_2013}{}. Accordingly more general models have been proposed. Some simply assume that the graphene and moir$\text{\'e}$ lattices are incommensurate\cite{ndiaye_structure_2008}{}. Others assume commensurability, yet without the constraint of a single moir$\text{\'e}$ beating within the moir$\text{\'e}$ unit cell. This situation is sketched in Fig.~\ref{fig:moire-sketch}c and accounts for experimental data obtained with gr/Ru and gr/Ir, for which four beatings were proposed in case of graphene’s zigzag rows aligning the metal’s dense-packed ones\cite{martoccia_graphene_2008,blanc_local_2012,jean_effect_2013,iannuzzi_moire_2013} and even more in graphene whose zigzag rows are $\sim30\degree$ rotated\cite{loginova_defects_2009}{}.

There are many ways the above assumption for commensurability can be fulfilled, as can be shown by considering strains and rotation of the graphene with respect to its substrate\cite{merino_strain-driven_2011}. For substrates exerting a weak bonding with graphene, rotations readily occur\cite{sutter_graphene_2009,loginova_defects_2009}{}. Strain, on the contrary, is more energetically costly in reason of the high mechanical stiffness of graphene\cite{lee_estimation_2012}{}. It appears that strains beyond few percents are not achievable in graphene synthesized on a substrate. Besides the isotropic strains considered to date, less symmetric deformations such as shear are also liable to be relevant\cite{hermann_periodic_2012}{}, but have only been marginally addressed in  the literature\cite{blanc_local_2012}{}.

Overall, a geometrical description of the full complexity of commensurate moir$\text{\'e}$ superlattices is missing. Here we extend the existing descriptions\cite{hermann_periodic_2012,zeller_what_2014}{}, relating the graphene, substrate, and moir$\text{\'e}$ lattices by a geometrical transformation, to the case of anisotropic deformations such as shear or uniaxial strain. This transformation is expressed within a matrix formalism and in an extension of the so-called Wood’s notation, which gives the angles formed between the unit cell vectors of graphene and the moir$\text{\'e}$ as well as the ratio between these vectors’ length. We use the latter notation to construct maps of the possible commensurate moir$\text{\'e}$ superlattices and to revisit previously published analysis of experimental observations. It shows supported graphene is subjected to strain levels far below what is usually assumed.

We apply this description to resolve the structure of the moir$\text{\'e}$ superlattices in graphene on multilayer graphene prepared on SiC, and in monolayer graphene on Re(0001) and Ir(111). For this purpose, we resort to scanning tunneling microscopy (STM) in both direct and reciprocal (Fourier) space, in the latter case achieving better than $0.1~\mathrm{pm}$ precision on the lattice parameter determination, owing distortion-less imaging with atomic resolution across several $10~\mathrm{nm}$ fields of view. We find rotated and sheared moir$\text{\'e}$ superlattices, which are well-reproduced with density functional theory (DFT) calculations. Some of these moir$\text{\'e}$s comprise several moir$\text{\'e}$ beatings in the case of metal substrates. Strikingly, commensurability between graphene and moir$\text{\'e}$ superlattices provides a fine description of even very large moiré supercells, comprising above $1,000$ carbon atoms.

\section*{Results}

\subsection*{General framework}

In most cases, supported graphene and its substrate do not share the same lattice parameter and/or graphene lies twisted by some angle with respect to its support. Assuming commensurability between the two lattices, a supercell can be defined which comprises the smallest integer numbers of unit cells of both graphene and the support. This supercell defines the moir$\text{\'e}$ superlattice. Formally, in a one-dimensional picture, the moir$\text{\'e}$ superlattice parameter $a_\mathrm{m}$ is an integer number times graphene’s ($a_\mathrm{gr}$) or the support’s ($a_\mathrm{s}$) lattice parameters: $a_\mathrm{m}=i\ a_\mathrm{gr}=m\ a_\mathrm{s}$, with $i$ and $m$ two coprime integers.

Still in one dimension, the reciprocal (Fourier) space unit vectors of the moir$\text{\'e}$ superlattice ($k_\mathrm{m}$), of graphene ($k_\mathrm{gr}$) and of the support ($k_\mathrm{s}$) hence fulfil $i\ k_\mathrm{m} = k_\mathrm{gr}$ and $m\ k_\mathrm{m} = k_\mathrm{s}$. We stress that these two equations constitute the general definition of a moir$\text{\'e}$ superlattice. On the contrary, the definition usually proposed in the literature, $k_\mathrm{m} = k_\mathrm{gr} - k_\mathrm{s}$, does not require commensurability. It can be obtained in the particular case of a commensurate system, with $i-m=1$, \textit{i.e.} with $i$ and $m$ two consecutive integer numbers. This particular case is sketched in Fig.~\ref{fig:moire-sketch}a. Figure~\ref{fig:moire-sketch}b shows a different situation with $i-m=2$. Strikingly, at first sight the two moir$\text{\'e}$s in Fig.~\ref{fig:moire-sketch}a,b are very similar. Indeed, at the middle of both linear ball models, the stacking of the carbon atoms onto the substrate ones is similar. In an analogy with optics, beatings between the two lattices seem to occur at the same location. Careful inspection however reveals that, for the $i-m=1$ moir$\text{\'e}$ (Fig.~\ref{fig:moire-sketch}a), the carbon atom sits exactly on top of the atom underneath, while for the $i-m=2$ moir$\text{\'e}$ (Fig.~\ref{fig:moire-sketch}b), the coincidence is only approximate. The difference is most often subtle in a scanning probe microscopy experiment\cite{iannuzzi_moire_2013} (similar graphene/support stackings yield similar signals), and usually overlooked, so the $i-m=2$ is generally (erroneously) described as a $i-m=1$ moir$\text{\'e}$. In fact it has a richer Fourier spectrum than the latter, as can be seen on Fig.~\ref{fig:moire-sketch}c. The fundamental Fourier harmonic of the $i-m=2$ moir$\text{\'e}$ is $k_\mathrm{m}=\nicefrac{(k_\mathrm{gr}-k_\mathrm{s})}{(i-m)}=\nicefrac{(k_\mathrm{gr}-k_\mathrm{s})}{2}$, and not $(k_\mathrm{gr}-k_\mathrm{s})$ as is the case for the $i-m=1$ moir$\text{\'e}$. The predominant intensity of the second harmonic ($k_\mathrm{gr}-k_\mathrm{s}$) translates nothing else than the close (but not exact) lattice coincidence observed at half the moir$\text{\'e}$ period (Fig.~\ref{fig:moire-sketch}b). The Fourier description of moir$\text{\'e}$s naturally makes the distinction between both, the $i-m=1$ moir$\text{\'e}$ containing only one beating, and the $i-m=2$ comprising two distinct ones.

The analysis of the moir$\text{\'e}$ superlattices presented below will be performed by expressing the moir$\text{\'e}$ superlattice unit vectors as function of those of the graphene and support unit cells. The analysis will also be expressed as function of elementary geometrical deformations, which we now introduce.

\begin{figure}[t]
\centering
\includegraphics{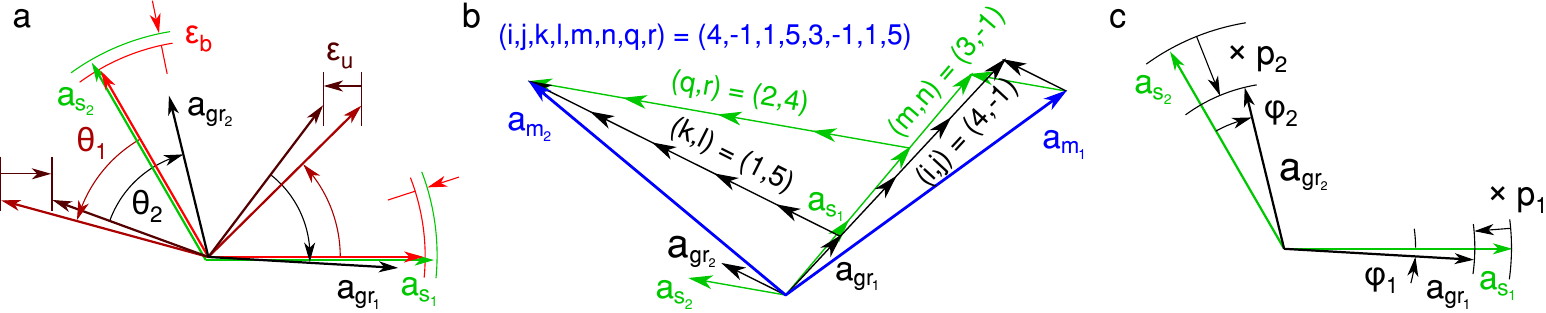}
\caption{\textbf{Structural interpretation of a moir$\text{\'e}$ superlattice}: (\textbf{a}) The transformation relating graphene lattice vectors ($\mathbf{a_{gr_1}}$,$\mathbf{a_{gr_2}}$) to those of its support ($\mathbf{a_{s_1}}$,$\mathbf{a_{s_2}}$) can be decomposed into four steps. (\textbf{1}) Graphene vectors are isotropically rescaled with respect to those of the support (light red). (\textbf{2}) Graphene is rotated with respect to its support (red), in order to determine the direction in which (\textbf{3}) a horizontal rescaling is applied (dark red). (\textbf{4}) A final rotation is applied (black). (\textbf{b}) The lattice vectors of the moir$\text{\'e}$ superlattice decompose into both graphene and support bases, giving $(i,j,k,l,m,n,q,r)=$ $(4,-1,1,5,3,-1,2,4)$. (\textbf{c}) Corresponding extended Wood's notation: $(p_1\ \mathrm{R}\varphi_1 \times p_2\ \mathrm{R}\varphi_2)$, where $p_1$ and $p_2$ are scaling factors, and $\varphi_1$ and $\varphi_2$ are rotation angles.}
\label{fig:struct_model}
\end{figure}

In the most general case, graphene is twisted, strained and sheared with respect to its substrate. The combination of all these contributions can be separated into four elementary geometrical transformations represented on Fig.~\ref{fig:struct_model}a: an isotropic rescaling (1), a directional rescaling (2 and 3), and a rotation (4). These transformations translate in mechanical terms as biaxial strain $\epsilon_{\mathrm{b}}$ (1), uniaxial strain in a given direction $\epsilon_{\mathrm{u}}$ (2 and 3), and a rotation (4) of the graphene layer. It can be noted that the so far overlooked shearing contribution is taken into account by combining a rotation, biaxial and uniaxial strains. The impact of such a combination on a moir$\text{\'e}$ has only been predicted\cite{hermann_periodic_2012}{}.

On the other hand, one can equivalently describe the graphene-substrate relation by explicitly writing the commensurate relation defining the moir$\text{\'e}$ superlattice. In order to account for its structural complexity in two dimensions, a set of eight integers $(i,j,k,l,m,n,q,r)$, which are determined through atomically-resolved microscopy, is then necessary (only four are needed to describe graphene maintaining the $D_{6h}$ symmetry, \textit{i.e.} when it is only strained biaxially and rotated):
\begin{equation} 
\left( \begin{array}{ccc}
\mathbf{a_{m_1}} \\
\mathbf{a_{m_2}} \end{array} \right) = 
M_\mathrm{gr}\left( \begin{array}{ccc}
\mathbf{a_{gr_1}} \\
\mathbf{a_{gr_2}} \end{array} \right) =
M_\mathrm{s}\left( \begin{array}{ccc}
\mathbf{a_{s_1}} \\
\mathbf{a_{s_2}} \end{array} \right)
\quad\mathrm{with}\quad
M_\mathrm{gr}=\left( \begin{array}{ccc}
i & j \\
k & l \end{array} \right) 
\mathrm{and}\ 
M_\mathrm{s}=\left( \begin{array}{ccc}
m & n \\
q & r \end{array} \right) \label{eqn:commensurability_DS}
\end{equation}
This translates into reciprocal space as:
\begin{equation}
\left( \begin{array}{ccc}
\mathbf{k_{gr_1}} \\
\mathbf{k_{gr_2}} \end{array} \right) = 
M_\mathrm{gr}^\mathrm{T}\left( \begin{array}{ccc}
\mathbf{k_{m_1}} \\
\mathbf{k_{m_2}} \end{array} \right)
\mathrm{\ and\ }
\left( \begin{array}{ccc}
\mathbf{k_{s_1}} \\
\mathbf{k_{s_2}} \end{array} \right) = 
M_\mathrm{s}^\mathrm{T}\left( \begin{array}{ccc}
\mathbf{k_{m_1}} \\
\mathbf{k_{m_2}} \end{array} \right)
\quad\mathrm{with}\quad
M_\mathrm{gr}^\mathrm{T}=\left( \begin{array}{ccc}
i & k \\
j & l \end{array} \right)
\mathrm{and}\ 
M_\mathrm{s}^\mathrm{T}=\left( \begin{array}{ccc}
m & q \\
n & r \end{array} \right)
\label{eqn:commensurability_RS}
\end{equation}
The $(i,j,k,l,m,n,q,r)$ integers used here correspond to the decomposition of the superstructure lattice vectors $\mathbf{a_{m_1}}$ and $\mathbf{a_{m_2}}$ into the basis formed by the graphene lattice vectors ($i,j,k,l$), and the supporting material lattice vectors ($m,n,q,r$), as sketched on Fig.~\ref{fig:struct_model}b. This decomposition is in practice performed more conveniently but equivalently in reciprocal space (Equation (\ref{eqn:commensurability_RS})).

By combining this description with that in terms of four geometrical transformations formally linking the graphene and support lattice vectors, one can relate the physical parameters describing how much graphene is strained and sheared to these eight integers. This relation is established in the Supplementary information (Equations~(S21a) and (S21b)) and is later used to quantify uniaxial and biaxial strains in graphene.

At this point we can generalize to the two-dimensional limit the concept of number of beatings $N$ in a moir$\text{\'e}$ cell. One can then define number of beatings $N_1$ and $N_2$ along $\mathbf{a_{m_1}}$ and $\mathbf{a_{m_2}}$ (see Supplementary information):
\begin{equation}
N_1 = \sqrt{(i-m)^2 + (j-n)^2 - (i-m)(j-n)}\ \mathrm{and}\ N_2 = \sqrt{(k-q)^2 + (l-r)^2 - (k-q)(l-r)}\label{eqn:moiron}
\end{equation}
The number of beatings $N$ within a moir$\text{\'e}$ cell is then simply given by the product $N=N_1N_2$.

Although using a set of eight integers is efficient to describe a moir$\text{\'e}$ superlattice, it is a relatively cumbersome notation that does not give an immediate picture of the structure. A clearer formulation of such sheared structures is then desirable. In the simple case of graphene experiencing only deformations preserving its pristine $D_{6h}$ symmetry, the Wood's notation circumvents this issue, describing the length and orientation of the superstructure lattice vectors compared to that of graphene or its supporting material. In the more general case addressed here, where the lattice vectors are allowed to vary in length and orientation independently as a result of shear and/or uniaxial strains, an extension of the Wood's notation is required, which we derive here. As depicted on Fig.~\ref{fig:struct_model}c, $\mathbf{a_{gr_1}}$ and $\mathbf{a_{gr_2}}$ are rescaled (resp. rotated) with respect to $\mathbf{a_{s_1}}$ and $\mathbf{a_{s_2}}$ by factors $p_1$ and $p_2$ (resp. angles $\varphi_1$ and $\varphi_2$). The extended Wood's notation reads as $(p_1\ \mathrm{R}\varphi_1 \times p_2\ \mathrm{R}\varphi_2)$. This notation gives the reader the ability to easily capture the graphene-substrate relation, and imagine how sheared it is by comparing $p_1$ and $p_2$, and $\varphi_1$ and $\varphi_2$. Once again, these quantities relate to the $(i,j,k,l,m,n,q,r)$ integers, as explained in the Supplementary information. The same can be done to relate the moir$\text{\'e}$ unit vectors to those of the support, as $(P_1\ \mathrm{R}\Phi_1 \times P_2\ \mathrm{R}\Phi_2)$, with:

\begin{center}
\begin{minipage}{.35\linewidth}
    \begin{subnumcases}{}
		p_1 = \sqrt{a^2 + b^2 - ab} \label{eqn:p_1}\\
		\varphi_1 = \mathrm{arctan} \left( \frac{b\sqrt{3}}{2a-b} \right) \label{eqn:alpha_1}\\
		p_2 = \sqrt{c^2 + d^2 - cd} \label{eqn:p_2}\\
		\varphi_2 = \mathrm{arctan} \left( \frac{c\sqrt{3}}{c-2d} \right) \label{eqn:alpha_2}
	\end{subnumcases}
\end{minipage}%
\begin{minipage}{.15\linewidth}
\centering
    $\mathrm{and}$
\end{minipage}%
\begin{minipage}{.35\linewidth}
    \begin{subnumcases}{}
		P_1 = \sqrt{m^2 + n^2 - mn} \label{eqn:P_1}\\
		\Phi_1 = \mathrm{arctan} \left( \frac{n\sqrt{3}}{2m-n} \right) \label{eqn:Phi_1}\\
		P_2 = \sqrt{q^2 + r^2 - qr} \label{eqn:P_2}\\
		\Phi_2 = \mathrm{arctan} \left( \frac{q\sqrt{3}}{q-2r} \right) \label{eqn:Phi_2}
	\end{subnumcases}
\end{minipage}
\end{center}
\noindent with $a=\frac{lm-jq}{il-jk}$, $b=\frac{ln-jr}{il-jk}$, $c=\frac{iq-km}{il-jk}$, $d=\frac{ir-kn}{il-jk}$.

\subsection*{Precision on the structure determination}

The geometrical developed so far proves necessary to properly interpret the refined twist angles and shearings observed in atomically-resolved microscopy images. The experimental uncertainty on the identification of the $(i,j,k,l,m,n,q,r)$ integers is here discussed to justify this necessity.

Quantitatively, the uncertainty on $(i,j,k,l)$ can be lowered by precisely determining $(\mathbf{k_{gr_1}},\mathbf{k_{gr_2}})$ and $(\mathbf{k_{m_1}},\mathbf{k_{m_2}})$. In practice, we measure the distance between the moir$\text{\'e}$ spots and the graphene spots in the Fourier transform image (each spot corresponding to a Fourier component), which are expected to be evenly separated. The sharpness of the spots is inversely proportional to the size of the atomically resolved image, and the number of spots increases with the contrast of the moir$\text{\'e}$ with respect to the atomic lattice. The former effect sets a precision in the determination of the spacing between two spots of $6\%$ in the case of gr/Ir, for which the image field of view is $\sim500~\mathrm{nm}^2$. The latter effect translates into an uncertainty as low as $\nicefrac{1}{\sqrt{50}}\times 6\%\sim 1\%$ in the case of gr/Ir (see Results below). Indeed, around the center of the reciprocal space, there are $\sim60$ Fourier components, which corresponds to $\sim50$ Fourier component spacings along one direction. The same is true around the graphene harmonics, so overall, in our example, the precision over $k_{\mathrm{m}}$ and $k_{\mathrm{gr}}$ is $\sim1\%$. For $(i,j,k,l)$, this precision translates through propagation of uncertainty into $2\%$.

The above described determination of $(i,j,k,l,m,n,q,r)$ is liable to put shears in evidence. At first thought, atomic resolution imaging can artificially produce sheared images. Such shears may result from imaging artefacts, for instance, in the case of scanning probe microscopy, thermal drift of the piezoelectric scanners or inequivalent calibration of these scanners along the two scan directions. However, these artefacts have no influence on the decomposition of $\mathbf{k_{gr_1}}$ and $\mathbf{k_{gr_2}}$ onto $\mathbf{k_{m_1}}$ and $\mathbf{k_{m_2}}$.

\subsection*{Twisted graphene bilayer}

First, the case of multilayer graphene on the C-rich $(000\bar{1})$ face (C-face) of a $\mathrm{4H}$-$\mathrm{SiC}$ sample is considered in Fig.~\ref{fig:gr-SiC}, where a $\sim 1.5$~nm beating is observed. In the present case, the relationship between the lattice vectors of the upper graphene layer and of the moir$\text{\'e}$ can be read on Fig.~\ref{fig:gr-SiC}b to deduce matrix $M_\mathrm{gr_{up}}^\mathrm{T}$. Here, this matrix indicates the coincidence of the graphene and moir$\text{\'e}$ lattices in reciprocal space. In direct space, this means that the beatings match the moir$\text{\'e}$, so $N=1$ (see General framework). For a $N=1$ moir$\text{\'e}$, $\mathbf{k_{m_1}}=\mathbf{k_{gr_1}^{up}}-\mathbf{k_{gr_1}^{low}}$, as the lower layer of graphene is the support material. From this, the matrix $M_\mathrm{gr_{low}}^\mathrm{T}$ between the lattice vectors of the lower graphene layer and of the moir$\text{\'e}$ is obtained. Transposing matrices $M_\mathrm{gr_{up}}^\mathrm{T}$ and $M_\mathrm{gr_{low}}^\mathrm{T}$ gives access to $M_\mathrm{gr_{up}}$ and $M_\mathrm{gr_{low}}$, which hold the decomposition of the moir$\text{\'e}$ unit vectors on the upper and lower graphene lattices in direct space:

\begin{figure}[t]
\centering
\includegraphics{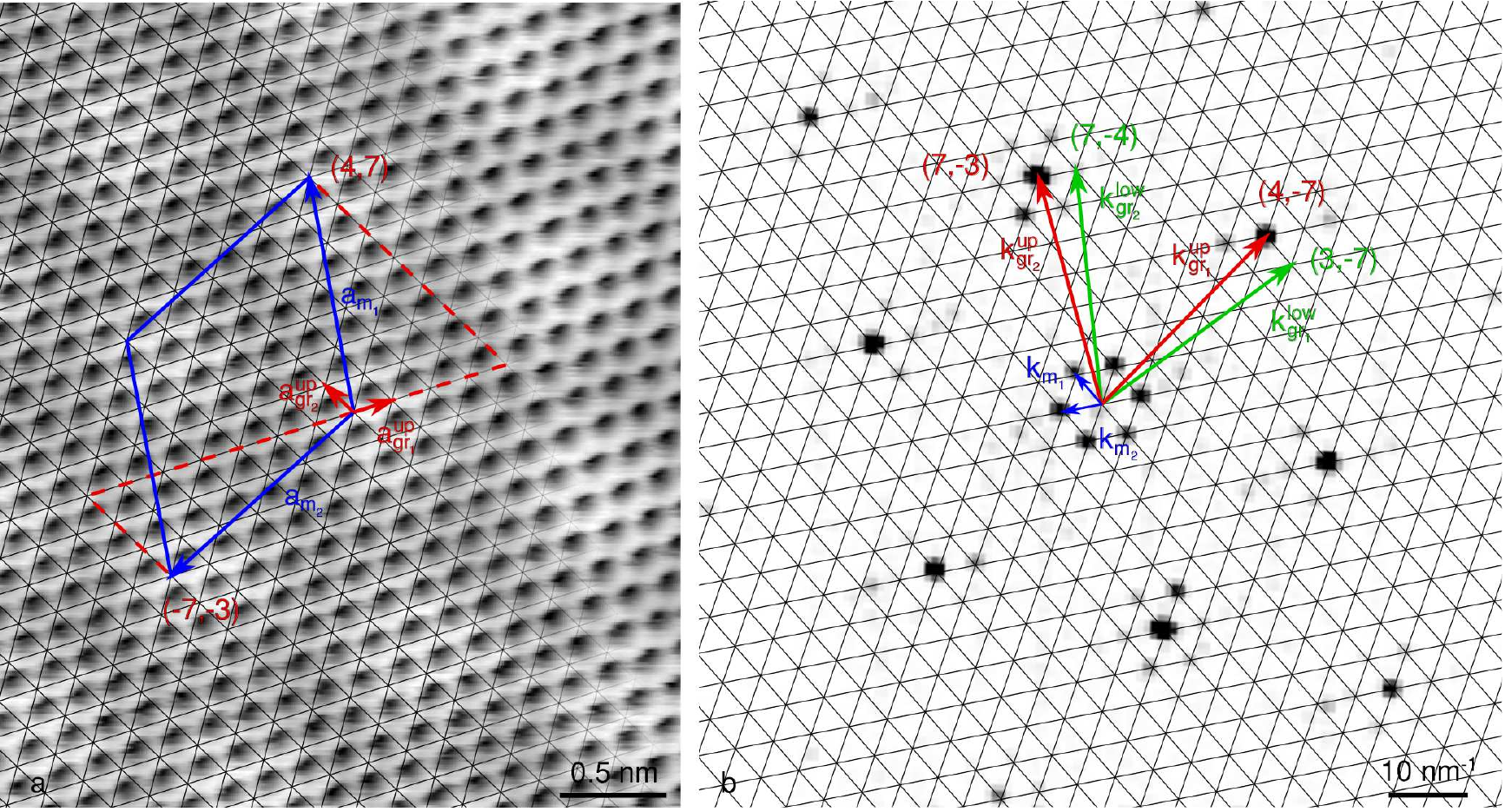}
\caption{\textbf{STM analysis of multilayer graphene on C-face SiC}: (\textbf{a}) (3.2~$\times$~3.8~nm$^2$) STM topograph ($I_\mathrm{tunnel}=10$~nA, $V_\mathrm{bias}=100$~mV) with emphasized upper graphene lattice (black), moir$\text{\'e}$ superlattice cell (blue rhombus) and lattice vectors of upper graphene and moir$\text{\'e}$ (red and blue respectively). (\textbf{b}) Corresponding FFT-image with emphasized moir$\text{\'e}$ reciprocal lattice (black) and lattice vectors of moir$\text{\'e}$ and upper and lower layers of graphene (blue, red and green respectively).}
\label{fig:gr-SiC}
\end{figure}

\begin{equation*} \label{eqn:gr-SiC}
M_\mathrm{gr_{up}}^\mathrm{T} = 
\left( \begin{array}{ccc}
4 & -7 \\
7 & -3 \end{array} \right)
\mathrm{and}\ 
M_\mathrm{gr_{low}}^\mathrm{T} = 
\left( \begin{array}{ccc}
3 & -7 \\
7 & -4 \end{array} \right)
\quad\mathrm{so}\quad
M_\mathrm{gr_{up}} = 
\left( \begin{array}{ccc}
4 & 7 \\
-7 & -3 \end{array} \right)
\mathrm{and}\ 
M_\mathrm{gr_{low}} = 
\left( \begin{array}{ccc}
3 & 7 \\
-7 & -4 \end{array} \right)
\end{equation*}

This commensurability relation gives a complete structural description, by decomposing the moir$\text{\'e}$ lattice vectors in the basis of each graphene layer, using the set of integers $(i,j,k,l,m,n,q,r)=(4,7,-7,3,3,7,-7,-4)$.
Using Equations~(\ref{eqn:p_1}), (\ref{eqn:alpha_1}), (\ref{eqn:p_2}) and (\ref{eqn:alpha_2}), such a structure can be analysed as two graphene layers sharing the same lattice parameter $a\mathrm{_{gr}^{up}}=a\mathrm{_{gr}^{low}}$ ($p_1=p_2=1$), and rotated by $\varphi=\varphi_1=\varphi_2=\mathrm{arctan} \left( \nicefrac{7\sqrt{3}}{73} \right) \sim 9.43\degree$ with respect to each other. This falls in the regime where the two graphene layers interact weakly, leading to Fermi velocity renormalization around the Dirac cones\cite{luican_single-layer_2011,ohta_evidence_2012,trambly_de_laissardiere_numerical_2012}{}.

\subsection*{Graphene on Re(0001)}

The case of multilayer graphene on C-face SiC has shown a situation where a moir$\text{\'e}$ superlattice is related to a single structural parameter: the twisting angle $\varphi$. On the contrary, graphene supported by a metallic surface can not only be twisted with respect to its substrate, but also strained, due to the lattice mismatch between the two. A full monolayer of graphene forms on Re(0001) through a self-limiting process\cite{miniussi_competition_2014}{}, and a $\sim 2.2$~nm period beating is found. These beatings were described as a $N=1$ moir$\text{\'e}$ superlattice with 10 graphene cells on 9 Re cells \cite{miniussi_thermal_2011}{}, or 8 graphene cells on 7 Re cells \cite{tonnoir_induced_2013}{}.

\begin{figure}
\centering
\includegraphics{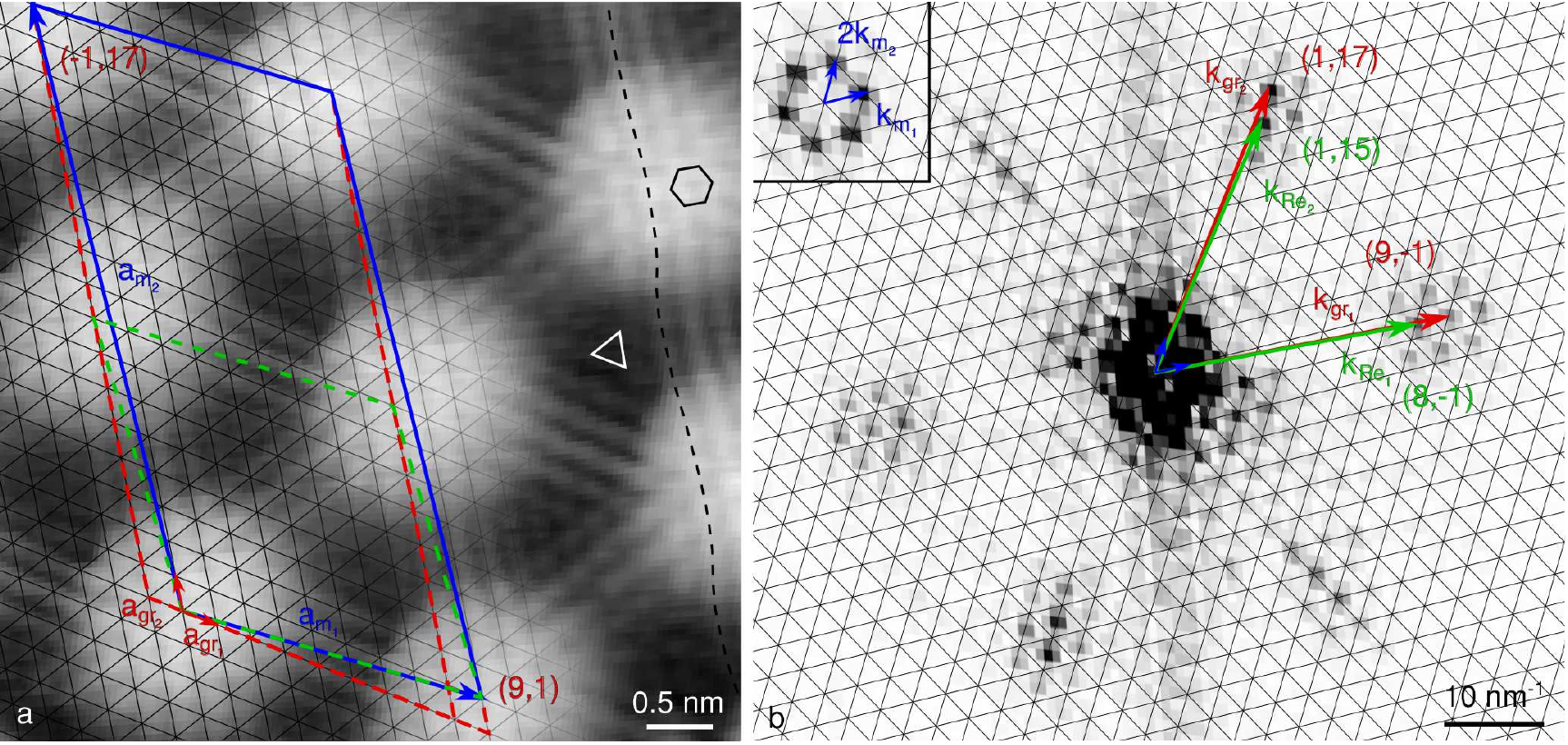}
\caption{\textbf{STM analysis of gr/Re}: (\textbf{a}) (5.6~$\times$~5.2~nm$^2$) STM topograph ($I_\mathrm{tunnel}=6$~nA, $V_\mathrm{bias}=30$~mV) with overlaid graphene lattice (black), and lattice vectors of graphene and $N=2$ superlattice (red and blue arrows respectively). Moir$\text{\'e}$ cell (blue full line) and its closest unsheared approximation with $N=1$ beating (green dashed line), with the coordinates of its corners in the graphene basis. The "odd-even" transition along lines of carbon atoms is also emphasized, as well as the either 6 or 3 C atoms observed in a moir$\text{\'e}$ hill or valley. (\textbf{b}) Corresponding FFT-image with emphasized moir$\text{\'e}$ reciprocal lattice (black) and lattice vectors of moir$\text{\'e}$, graphene and Re (blue, red and green respectively). Inset shows the $\mathbf{k_{gr}}-\mathbf{k_{Re}}$ harmonics surrounding the center of the FFT-image with improved contrast.}
\label{fig:gr-Re}
\end{figure}

A direct analysis of the STM topograph along the same lines as for Fig.~\ref{fig:gr-SiC} is here challenging. Figure~\ref{fig:gr-Re}a highlights two additional phenomena, which have been little discussed to our knowledge in the context of graphene on metals\cite{marchini_scanning_2007}{}.

First, depending on the position within the beatings, the apparent height accessed by STM shows a varying number of visible C: in a valley, only 3 atoms out of a 6-C ring are seen, whereas on a hill, all 6 are observed. This is due to the sites occupied by the C atoms on the terminal metallic layer, which fall into three typical configurations: a C on top of a metal atom (atop), or on top of a hollow site. Two kinds of hollow site can be distinguished, depending on the presence (hcp) or absence (fcc) of another metal atom of the second terminal metallic layer below the hollow site. In a valley, the sites occupied by the C atoms are either atop and hcp, or atop and fcc. The overlap of the $p_\mathrm{z}$-like orbital of a C atom in atop position with the $d$-like orbitals of the underlying metal atom is then maximal. Consequently, the local electronic density of states is modified, making it appear low in STM \cite{wang_chemical_2008}{}. This explains why only half the C atoms appear as protrusions in a moir$\text{\'e}$ valley, while all of them can be identified on top of a moir$\text{\'e}$ hill.

Second, the apparent atomic rows of C oscillate with the same periodicity as the beatings. This phenomenon has been reported and discussed in the case of gr/Ru \cite{marchini_scanning_2007}{}, but is known since the 1990s as the "odd-even transition" in the case of graphite \cite{nysten_afm-stm_1993,miyake_giant_1998,osing_bulk_1998}{}. Its origin is well illustrated in the case of the two distinctive moir$\text{\'e}$ valleys. Indeed, they differ only in the site of the remaining visible C atom: hcp or fcc. Depending on whether the site is hcp or fcc, the corresponding C atom belongs to sub-lattice A or B of graphene. As a consequence, when moving from one beating to the other, the C atoms that are observed switch continuously from one sub-lattice to the other. Along a row of C atoms, this induces an apparent oscillation of the row. Therefore, these two effects are related to a modulation of the electronic density of states on the two sub-lattices of graphene, which is correlated with the moir$\text{\'e}$ periodicity.

Using DFT calculations, these two effects have been reproduced in the case of a sheared and twisted $N=1$ moir$\text{\'e}$ superlattice of gr/Re, comprising a sufficiently small number of atoms to be treated numerically. This moir$\text{\'e}$ is characterized by the set $(i,j,k,l,m,n,q,r)=(9,3,-2,7,8,3,-2,6)$. On Fig.~S2b, one can see that the moir$\text{\'e}$ reproduces the two anomalies described above. Only one C sub-lattice is observed in each moir$\text{\'e}$ valley. Within the unit cell, this causes an effective oscillation of the atomic C row, which is actually related to the varying contribution of each sub-lattice to the electronic density, as can be checked on Fig.~S2a.

On Fig.~S2b, it can be noted that the hills of the beating are not circular, but appear rather elliptical. This is attributed to the small shearing that graphene undergoes in this superstructure, whose effect is enhanced on the moir$\text{\'e}$. Similar non-circular hills can be observed on Fig.~\ref{fig:gr-Re}a, which is another clue that indicates graphene structure is sheared on this STM topograph.

Figures~\ref{fig:gr-Re}a-b display an analysis taking the two STM electronic effects into account. The FFT image is analysed similarly to Fig.~\ref{fig:gr-SiC}b, although the situation is different. Indeed, in two directions, the graphene spots do not superimpose with the extrapolated reciprocal space lattice paved with the $\mathbf{k_{gr_1}}-\mathbf{k_{Re_1}}$ and $\mathbf{k_{gr_2}}-\mathbf{k_{Re_2}}$ vectors. This means that the moir$\text{\'e}$ is not a $N=1$ superlattice (cf. Fig.~\ref{fig:moire-sketch}b). Moreover, the positions of the graphene spots with respect to the moir$\text{\'e}$ reciprocal network vary for the three main directions. Consequently, based on the reciprocal space analysis, the moir$\text{\'e}$ structure considered here is sheared. The commensurability relation of this structure reads as:

\begin{equation*} \label{eqn:gr/Re}
M_\mathrm{gr}^\mathrm{T} = 
\left( \begin{array}{ccc}
9 & -1 \\
1 & 17 \end{array} \right)
\mathrm{and}\ 
M_\mathrm{Re}^\mathrm{T} = 
\left( \begin{array}{ccc}
8 & -1 \\
1 & 15 \end{array} \right)
\quad\mathrm{so}\quad
M_\mathrm{gr} = 
\left( \begin{array}{ccc}
9 & 1 \\
-1 & 17 \end{array} \right)
\mathrm{and}\ 
M_\mathrm{Re} = 
\left( \begin{array}{ccc}
8 & 1 \\
-1 & 15 \end{array} \right)
\end{equation*}

The corresponding set of integers therefore is $(i,j,k,l,m,n,q,r)=(9,1,-1,17,8,1,-1,15)$. As a signature of the anisotropy, the moir$\text{\'e}$ cell contains a different number of beatings $N_1=1$ and $N_2=2$ in each of its main directions, as can be deduced from Equation~(\ref{eqn:moiron}). This analysis is displayed in direct space on top of the original STM topograph on Fig.~\ref{fig:gr-Re}a, where the superstructure lattice vectors are explicitly decomposed on the graphene lattice.

To get a more simple grasp of this structure, the moir$\text{\'e}$ can be described using the $(P_1\ \mathrm{R}\Phi_1 \times P_2\ \mathrm{R}\Phi_2)$ extended Wood's notation, with $P_1=\sqrt{8^2+1^2-1\times8}\sim7.55$, $P_2=\sqrt{(-1)^2+15^2-(-1)\times15}\sim15.52$, $\Phi_1=\arctan\left(\nicefrac{\sqrt{3}}{15}\right)\sim6.59\degree$, and $\Phi_2=\arctan\left(\nicefrac{\sqrt{3}}{31}\right)\sim3.20\degree$, as deduced from Equations~(\ref{eqn:P_1}), (\ref{eqn:Phi_1}), (\ref{eqn:P_2}) and (\ref{eqn:Phi_2}). This notation makes clear the twice larger size of the moir$\text{\'e}$ compared to a $N=1$ superlattice, comprising 308 carbon atoms, as well as a sizeable shear. The corresponding shear of the graphene lattice is obvious in the corresponding extended Wood's notation $(p_1\ \mathrm{R}\varphi_1 \times p_2\ \mathrm{R}\varphi_2)$. Using Equations~(\ref{eqn:p_1}), (\ref{eqn:alpha_1}), (\ref{eqn:p_2}) and (\ref{eqn:alpha_2}), one gets $p_1=\nicefrac{\sqrt{137^2+2^2-2\times137}}{154}\sim 0.883$, $p_2=\nicefrac{\sqrt{(-1)^2+136^2+1\times136}}{154}\sim 0.886$, $\varphi_1=\mathrm{arctan}\left(\nicefrac{\sqrt{3}}{136}\right)\sim 0.73\degree$ and $\varphi_2=\mathrm{arctan}\left(\nicefrac{\sqrt{3}}{271}\right)\sim 0.36\degree$. This is summarized as $(0.883\ \mathrm{R}0.73\degree \times 0.886\ \mathrm{R}0.36\degree)$. This structure is close but different from the previously reported assignment of a $(7\ \times 7)\mathrm{R}0\degree$ $N=1$ moir$\text{\'e}$ (Ref.~\citen{tonnoir_induced_2013}).

Overall, the structure is a superlattice both sheared and twisted with $N_1=1$ beating in one direction and $N_2=2$ in the other, giving rise to $N=2$ beatings in the moir$\text{\'e}$ cell. This is significantly more complex than the $N=1$ twisted superlattices discussed in many reports, and even than the $N=4$ untwisted superlattices reported in gr/Ir\cite{blanc_strains_2013} and gr/Ru\cite{martoccia_graphene_2008}{}, or than a solely sheared superlattice\cite{hermann_periodic_2012}{}.

A more physical description of such a structure is given by comparing the graphene overlayer with its HOPG counterpart, and decomposing the strain in terms of a uniaxial and a biaxial contributions. Using Equations~(S21a) and (S21b) in the case of gr/Re, graphene is biaxally compressed by $\epsilon_{\mathrm{b}}\sim-0.14\%$ and uniaxially compressed by $\epsilon_{\mathrm{u}}\sim-0.84\%$. This shows a moir$\text{\'e}$ is actually related to a non-trivial distortion of the graphene lattice.

\subsection*{Graphene on Ir(111)}

\begin{figure}[t]
\centering
\includegraphics{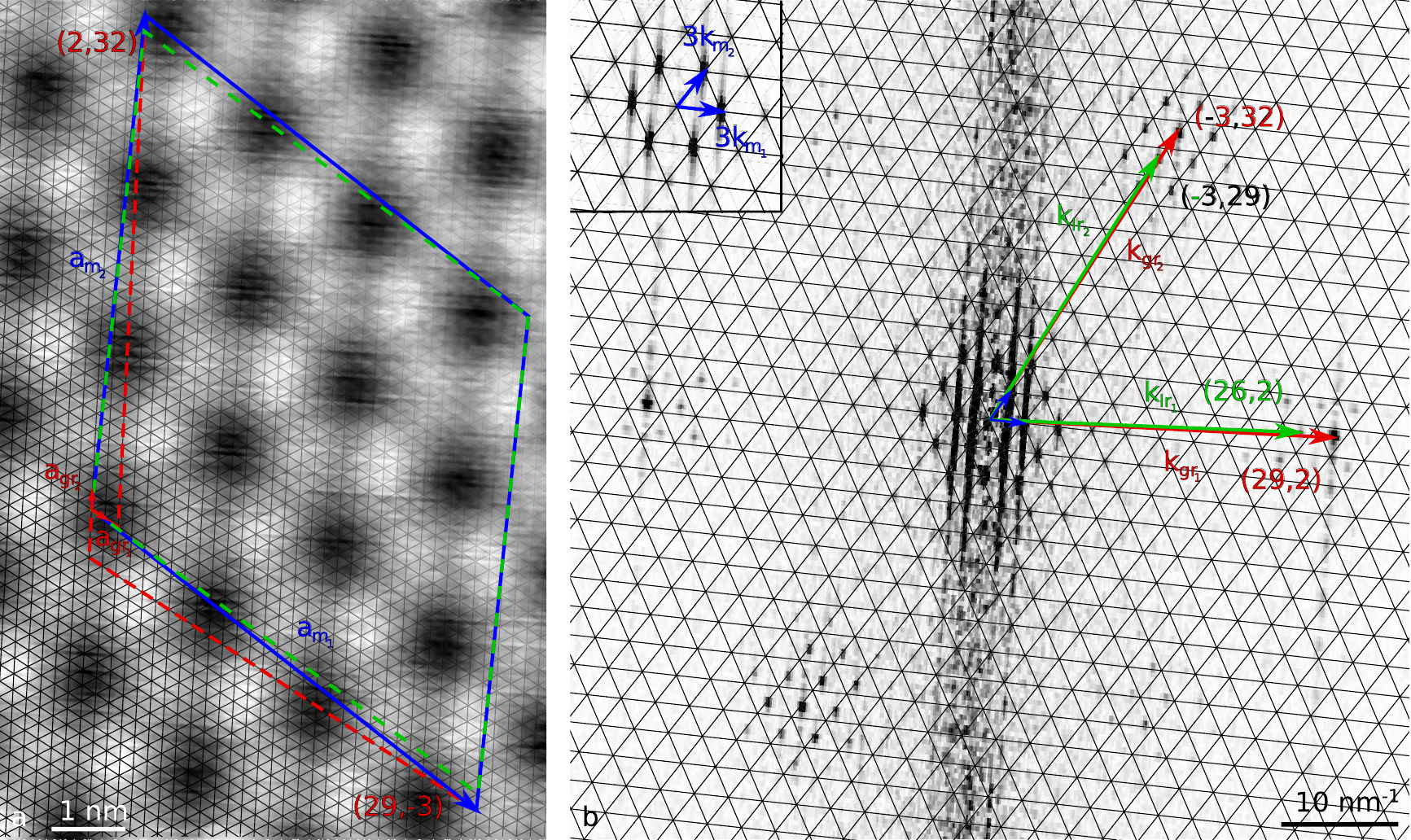}
\caption{\textbf{STM analysis of gr/Ir}: (\textbf{a}) (13.5~$\times$~7.4~nm$^2$) STM topograph ($I_\mathrm{tunnel}=20$~nA, $V_\mathrm{bias}=60$~mV) with highlighted graphene lattice (black), and lattice vectors of graphene and $N=9$ moir$\text{\'e}$ (red and blue arrows respectively). Moir$\text{\'e}$ cell (blue line) with the coordinates of its corners in the graphene basis, and its closest unsheared approximation (green dashed line). It should be noted the contrast is inverted compared to Fig.~\ref{fig:gr-Re}, so hills appear as dark. (\textbf{b}) FFT-image obtained from a 15.6~$\times$~30~nm$^2$ STM topograph, and overlaid with the lattice paved with $\mathbf{k_{gr}}-\mathbf{k_{Ir}}$ vectors, and lattice vectors of moir$\text{\'e}$, graphene and Ir (blue, red and green respectively). Inset shows moir$\text{\'e}$ spots surrounding the center of the FFT-image with improved contrast.}
\label{fig:gr-Ir}
\end{figure}

The anisotropy of the graphene and moir$\text{\'e}$ lattices is also encountered when the graphene-substrate interaction is much weaker, \textit{e.g.} gr/Ir. Similarly to gr/Re, the FFT-image of Fig.~\ref{fig:gr-Ir}b shows the graphene spots do not superimpose with the extrapolated reciprocal lattice paved with $\mathbf{k_{gr_1}}-\mathbf{k_{Ir_1}}$ and $\mathbf{k_{gr_2}}-\mathbf{k_{Ir_2}}$, which means the moir$\text{\'e}$ comprises more than a single beating ($N>1$). In addition, the position of the graphene spots with respect to the moir$\text{\'e}$ reciprocal lattice is not the same in each main direction, which means the structure is sheared. Actually, along the close-to-horizontal direction in reciprocal space (center-right in Fig.~\ref{fig:gr-Ir}b), the set of harmonics around $\mathbf{k_{gr_1}}$ are for instance found right at the center of mass of the triangles defined by the extrapolated lattice. On the contrary, for the second direction (top-right in Fig.~\ref{fig:gr-Ir}b), the set of harmonics around $\mathbf{k_{gr_2}}$ lie in between two nodes of the extrapolated reciprocal lattice. This translates into the commensurability relation as:
\begin{equation*} \label{eqn:gr/Ir}
M_\mathrm{gr}^\mathrm{T} = 
\left( \begin{array}{ccc}
29 & 2 \\
-3 & 32 \end{array} \right)
\mathrm{and}\ 
M_\mathrm{Ir}^\mathrm{T} = 
\left( \begin{array}{ccc}
26 & 2 \\
-3 & 29 \end{array} \right)
\quad\mathrm{so}\quad
M_\mathrm{gr} = 
\left( \begin{array}{ccc}
29 & -3 \\
2 & 32 \end{array} \right)
\mathrm{and}\ 
M_\mathrm{Ir} = 
\left( \begin{array}{ccc}
26 & -3 \\
2 & 29 \end{array} \right)
\end{equation*}

This description of the superlattice can be summarized with $(i,j,k,l,m,n,q,r)=(29,-3,2,32,26,-3,2,29)$, as interpreted in Fig.~\ref{fig:gr-Ir}a. Such a moir$\text{\'e}$ comprises three beatings in each direction (Equation~(\ref{eqn:moiron})), in total $1,868$ carbon atoms. In the extended Wood's notation, this superlattice is described with $(P_1\ \mathrm{R}\Phi_1 \times P_2\ \mathrm{R}\Phi_2)$, with $P_1=\sqrt{26^2+(-3)^2+26\times3}{\sim27.62}$, $P_2=\sqrt{2^2+29^2-2\times29}\sim28.05$, $\Phi_1=\mathrm{arctan}\left(\nicefrac{-3\sqrt{3}}{55}\right)\sim-5.40\degree$, and $\Phi_2=\mathrm{arctan}\left(\nicefrac{-\sqrt{3}}{28}\right)\sim-3.54\degree$, as deduced from Equations~(\ref{eqn:P_1}), (\ref{eqn:Phi_1}), (\ref{eqn:P_2}) and (\ref{eqn:Phi_2}). This is very close but still different from the so-called incommensurate $(9.32\ \times 9.32)~\mathrm{R}0\degree$ structure\cite{ndiaye_structure_2008}{}. The graphene structure is similarly described with $(p_1\ \mathrm{R}\varphi_1 \times p_2\ \mathrm{R}\varphi_2)$, with $p_1=\nicefrac{\sqrt{838^2+(-9)^2-838\times(-9)}}{934}\sim0.902$, $p_2=\nicefrac{\sqrt{6^2+847^2-6\times847}}{934}\sim0.904$, $\varphi_1=\mathrm{arctan}\left(\nicefrac{-9\sqrt{3}}{1685}\right)\sim-0.53\degree$, and $\varphi_2=\mathrm{arctan}\left(\nicefrac{-3\sqrt{3}}{844}\right)\sim-0.35\degree$, as deduced from Equations~(\ref{eqn:p_1}), (\ref{eqn:alpha_1}), (\ref{eqn:p_2}) and (\ref{eqn:alpha_2}). These values are in excellent agreement with the $0.903$ ratio recently measured by means of surface X-ray scattering\cite{jean_topography_2015}{}.

Using Equations~(S21a) and (S21b), this shearing translates into a combination of biaxial compression $\epsilon_{\mathrm{b}}\sim-0.29\%$ and uniaxial compression $\epsilon_{\mathrm{u}}\sim-0.41\%$ (expressed using HOPG as a reference for unstrained graphene). Shear and strain of such extents have already been reported before\cite{blanc_local_2012}{}, but no quantitative analysis was provided.

\section*{Discussion}

\begin{figure}
\centering
\includegraphics{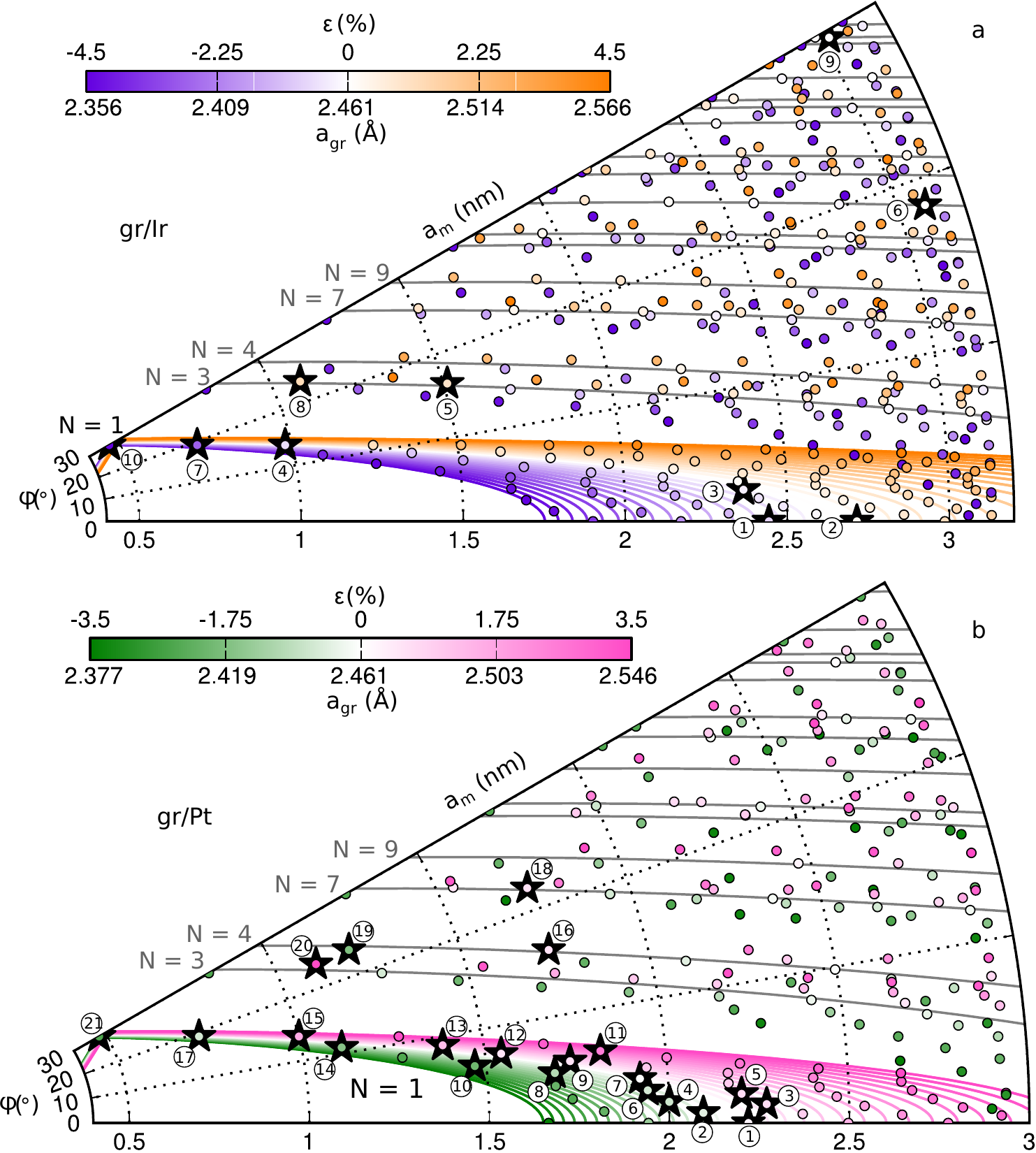}
\caption{\textbf{Moir$\text{\'e}$ lattice constant $a_\mathrm{m}$ versus angle $\varphi$ between graphene and its support}. Each point corresponds to a commensurate superlattice of given $a_\mathrm{m}$ and $\varphi$, with its colour indicating the strain level of graphene. The full lines indicate the superlattices containing $N$ beatings within the moir$\text{\'e}$ cell. The coloured ones add a strain information, and for clarity are only shown for the $N=1$ case. (\textbf{a}) gr/Ir, (\textbf{b}) gr/Pt (with lattice parameters $a_{\mathrm{Ir}}=2.7147~\text{\AA}$ and $a_{\mathrm{Pt}}=2.7744~\text{\AA}$, see Ref.~\citen{lide_crc_1994}). Black stars index reported unsheared structures (see Supplementary information for detailed references).}
\label{fig:moire1}
\end{figure}

Three support lattices have been considered so far, revealing that a moir$\text{\'e}$ structure can be rotated, strained and sheared. It also demonstrates that moir$\text{\'e}$ superlattices comprising more than one beating are commonly encountered. Three equivalent ways have been presented to describe moir$\text{\'e}$ superlattices with ease:
\begin{itemize}
\itemsep0em
\item Using an extended Wood's notation for a pictorial description using two scaling factors and two angles,
\item In more physical terms with rotation angles, and uniaxial and biaxial strains,
\item With eight integers that decompose independently the two moir$\text{\'e}$ lattice vectors onto those of graphene and of its support.
\end{itemize}
The latter allows for enumerating all the possible structures by combining every possible value for each integer. The system can also be treated by addressing two independent directions separately, \textit{i.e.} by considering two sets of four integers, $(i,j,m,n)$ and $(k,l,q,r)$, which obey the same equations. Below, we introduce these equations in the case of $(i,j,m,n)$ from Equations~(\ref{eqn:commensurability_DS}) and (\ref{eqn:commensurability_RS}):
\begin{equation} \label{eqn:gr/supp}
M_\mathrm{gr} = 
\left( \begin{array}{ccc}
i & j \\
-j & i-j \end{array} \right)
\mathrm{,\ }
M_\mathrm{s} = 
\left( \begin{array}{ccc}
m & n \\
-n & m-n \end{array} \right)
\quad\mathrm{and}\quad
M_\mathrm{gr}^\mathrm{T} = 
\left( \begin{array}{ccc}
i & -j \\
j & i-j \end{array} \right)
\mathrm{,\ }
M_\mathrm{s}^\mathrm{T} = 
\left( \begin{array}{ccc}
m & -n \\
n & m-n \end{array} \right)
\end{equation}
Out of these two equivalent equations, and using only the support lattice constant $a_{\mathrm{s}}$, one can express the moir$\text{\'e}$ lattice constant $a_\mathrm{m}$, the biaxial strain $\epsilon$ (assuming the lattice parameter of HOPG as a zero-strain situation), and the twist angle $\varphi$ between graphene and its support (see Supplementary information):
\begin{equation} \label{eqn:gr-supp_iso}
\begin{cases}
a_\mathrm{m} = a_{\mathrm{s}} \sqrt{m^2+n^2-mn} \\
\epsilon = \frac{a_{\mathrm{gr}} - a_{\mathrm{HOPG}}}{a_{\mathrm{HOPG}}} = \frac{a_{\mathrm{s}}}{a_{\mathrm{HOPG}}} \sqrt{\frac{m^2+n^2-mn}{i^2+j^2-ij}}-1 \\
\varphi = \mathrm{arctan} \left( \frac{(in-jm)\sqrt{3}}{2(im+jn)-(in+jm)} \right)
\end{cases}
\end{equation}
The link with their Wood's notation $(p\times p)\mathrm{R}\varphi$ can be established straightforwardly (see Supplementary information) as:
\begin{equation} \label{eqn:gr-supp_woods}
p = \frac{a_\mathrm{gr}}{a_\mathrm{s}}=\sqrt{\frac{m^2+n^2-mn}{i^2+j^2-ij}}\ \mathrm{and}\ \varphi = (\mathbf{a_{s}},\mathbf{a_{gr}}) = \mathrm{arctan} \left( \frac{(in-jm)\sqrt{3}}{2(im+jn)-(in+jm)} \right)
\end{equation}
With increasing values of $(i,j,m,n)$, it is then possible to enumerate every possible structure. Using a limited set of parameters such as those defined by Equation~(\ref{eqn:gr-supp_woods}) allows for a graphical representation of the strain of every possible moir$\text{\'e}$ superlattice in a given direction. Figure~\ref{fig:moire1} gives this representation in the case of graphene on dense-packed surfaces of various metals (Re, Ir, Pt). Figure~\ref{fig:moire1} also displays parametrized curves accounting for a definition of a moir$\text{\'e}$ superlattice with no assumption on commensurability, corresponding to the one-dimensional formula $Nk_m=k_{gr}-k_s$. The representation of these curves is given by:

\begin{equation}
a_\mathrm{m} = a_{\mathrm{s}} \sqrt{\frac{N}{1-2p\ \mathrm{cos}\varphi+p^2}} \label{eqn:curve_moiron}
\end{equation}
Where $p=\dfrac{a_{\mathrm{gr}}}{a_{\mathrm{s}}}=\dfrac{a_{\mathrm{HOPG}}}{a_{\mathrm{s}}}\left(\epsilon+1\right)$ and $N$ the number of beatings given by Equation~(\ref{eqn:moiron}) with $N_1=N_2$.

This series of parametrized curves highlights moir$\text{\'e}$ superlattices with increasing numbers of beatings $N$ (see Supplementary information). The above-described results, as well as data extracted from the literature in the case of gr/Pt and gr/Ir, are displayed on Fig.~\ref{fig:moire1}.


In the case of gr/Pt (Fig.~\ref{fig:moire1}b), the interpretation in terms of sub-3 nm period superlattices corresponds to suspiciously high strains for a system with such a weak interaction between graphene and the metal. For instance, the phases of gr/Pt indexed as 11, 19, 20 and 21 on Fig.~\ref{fig:moire1}b have been interpreted as moir$\text{\'e}$ superlattices with respectively $\epsilon=2.51\%$ (11), $\epsilon=-1.73\%$ (19) and $\epsilon=3.45\%$ (20), and $\epsilon=-2.38\%$ (21). Higher number of beatings are in fact probable for such structures. Such a high number of beatings was determined in the case of the so-called R30 phase of gr/Ir\cite{loginova_defects_2009}{}. A combined micro-spot low energy electron diffraction ($\mu$-LEED) and STM study showed that within a moir$\text{\'e}$ unit cell of $\sim3.02~\mathrm{nm}$ lattice parameter, $N=37$ beatings separated by $\sim0.47~\mathrm{nm}$ occur ($N_1=N_2=\sqrt{37}$). This $N=37$ moir$\text{\'e}$ is described with $(i,j,m,n)=(14,9,12,2)$ (indexed as 9 on Fig.~\ref{fig:moire1}a), which corresponds to $\epsilon=-0.04\%$. This moir$\text{\'e}$ was also described as a $N=1$ moir$\text{\'e}$\cite{meng_multi-oriented_2012} with $(i,j,m,n)=(2,0,2,1)$ (indexed as 10 on Fig.~\ref{fig:moire1}a), for which $\epsilon=-4.48\%$, which is questionable. Similarly, the so-called R18.5 of gr/Ir was interpreted as either $(i,j,m,n)=(13,1,13,5)$, $\varepsilon=-0.02\%$ (Ref.~\citen{loginova_defects_2009}), or $(i,j,m,n)=(3,0,3,1)$, $\varepsilon=-2.73\%$ (Ref.~\citen{zeller_scalable_2012}), respectively labelled as 6 and 7 on Fig.~\ref{fig:moire1}a.

The analysis performed here demonstrates the rich variety of moir$\text{\'e}$ superlattices to be expected for graphene on a substrate, well beyond the simple case of $N=1$ unsheared cases. Although many structures are possible from the geometrical point of view, few of them have actually been reported in the literature. This state of fact can be interpreted in two different ways: either differentiating some very similar structures is not possible due to too limited space resolution, or only the possibility has not been considered, or only a few of them are stable enough to actually exist.

Gr/Ir and gr/Pt are typical of the first situation. Numerous moir$\text{\'e}$ phases have been reported for them, as shown on Fig.~\ref{fig:moire1}a-b. The majority of them is identified as $N=1$ moir$\text{\'e}$ superlattices, nevertheless, this description appears sometimes unrealistic. For gr/Re, like gr/Ru and gr/Ni, graphene tends to align its zigzag rows to the metal's close-packed rows ($\varphi\sim0\degree$), even in growth conditions quite far from thermodynamic equilibrium. Presumably, the strong bonds of covalent character between carbon and metal atoms inside the growing flake are not readily broken, as would be required for twisting.

Although large-angle graphene twists are almost prohibited for gr/Re, slightly twisted graphene phases of gr/Re coexist. These numerous very similar structures can be assumed to be local minima in the energy landscape of gr/Re. Their coexistence then implies a high activation energy between each of them, so the formation of a large-scale uniform graphene phase is kinetically limited. In other words, graphene needs to be heated to high enough temperature to rearrange into the most stable phase of gr/Re. However, at high temperature, graphene growth competes with bulk dissolution and carbide formation, so the growth is performed by annealing cycles\cite{miniussi_competition_2014}{}. Over each cycle, graphene's crystallinity progressively improves, which supports this simple kinetic scenario. To go further, one can compare this situation with that of gr/Ru, where domains slightly rotated around $\varphi\sim0\degree$ can be grown, as observed in STM\cite{natterer_optimizing_2012} and $\mu$-LEED\cite{man_small-angle_2011}{}. By tuning the growth to higher temperature, large domains of one specific structure tend to form\cite{natterer_optimizing_2012,pan_highly_2009}{}, which has been analysed as a $N=4$ ($N_1=N_2=2$) superlattice ($(i,j,m,n)=(25,0,23,0)$) using surface x-ray diffraction\cite{martoccia_graphene_2008}{}. Such similar behaviours may lead to the conclusion that the mechanism presented here is common to every system where graphene is in strong interaction with its substrate.

Graphene on C-face SiC grows with rotational disorder between the adjacent graphene layers\cite{hass_why_2008}{}, so the terminal layers exhibit many possible twisted phases\cite{varchon_rotational_2008}{}. Even though all kinds of twists are encountered in experiments, it seems that certain twist angles are preferential. We surmise that these twist angles correspond to commensurate moir$\text{\'e}$ superlattices such as the one that we report. Since both graphene layers share the same lattice parameter, the situation can be depicted with two integers $(i,j)$, such that $(i,j,m,n)=(i,j,-j,i-j)$. For instance, $(i,j)=(4,7)$ in the present work, and $(i,j)=(4,1)$ in Ref.~\citen{hass_why_2008}{}. We note that the observation, with diffraction techniques, of a continuum of twist angles (e.g. see Ref.~\citen{hass_why_2008}{}) does not necessarily imply that the twist angle can take random values. Indeed, the existence of a multitude of commensurate superlattices discretely spanning the $0-60\degree$ twist range could as well account for the observation due to the finite size of the diffraction spot (set by the domain size or the instrumental resolution) that they yield.

In conclusion, different supported graphene systems have been studied with STM. A consistent analysis of moir$\text{\'e}$ superlattices involving both direct and FFT STM images has been presented, consistent with DFT calculations. A spatial precision of a tenth of $1~\mathrm{pm}$ is achieved, revealing that graphene lying on a substrate is actually twisted, strained and sheared, which breaks its rotational symmetry. A geometrical model enables to classify all moir$\text{\'e}$ superlattices. This model gives a global picture assuming commensurability between graphene and its substrate (and consequently between graphene or the substrate, and the moir$\text{\'e}$), yielding various numbers of beatings. While a very large number of structures is possible, only a few have actually been reported. In the case of strong graphene-substrate interaction, it is unlikely that all predicted superlattices are discovered, since for instance phases corresponding to a substantial rotation of graphene with respect to the substrate do not tend to form. For low interaction graphene-substrate systems, the complexity of the moir$\text{\'e}$ superlattices has been undetected or overlooked, leading to possibly simplified interpretations. We anticipate that moir$\text{\'e}$ superlattices with $N>1$ number of beatings will produce rich electronic modulations in graphene.

\section*{Methods}

\textbf{Preparation of multilayer gr/SiC}. Graphene has been grown on undoped double-polished $\mathrm{4H}$-$\mathrm{SiC}(000\bar{1})$, purchased from Novasic and cut into $5\times5~\mathrm{mm}^2$ pieces. The growth has been performed in a RF-furnace following the recipe in Ref.~\citen{emtsev_towards_2009}. SiC surface was first cleaned in $\mathrm{H}_2$ and Ar atmosphere at $1,560~\degree\mathrm{C}$, and subsequently annealed in Ar atmosphere at the same temperature.

\noindent \textbf{Preparation of gr/Re}. Re single crystal cut in the (0001) surface purchased from Surface Preparation Laboratory was cleaned in a ultrahigh vacuum (UHV) chamber (base pressure~$\sim10^{-10}~\mathrm{mbar}$) by cycles of Ar$^+$ ion bombardment at $2~\mathrm{keV}$ at $750~\degree\mathrm{C}$ and subsequent annealing at $\sim1,300~\degree\mathrm{C}$.
The gr/Re was prepared following the recipe presented in Ref.~\citen{miniussi_competition_2014}, by saturating the Re(0001) surface with $\mathrm{C_2H_4}$ at room temperature (introduced with a $3\cdot10^{-8}~\mathrm{mbar}$ pressure), and two subsequent cycles of flash-annealing/cooling at $750~\degree\mathrm{C}$ with a $5\cdot10^{-7}\mathrm{mbar}$ $\mathrm{C_2H_4}$ pressure.

\noindent \textbf{Preparation of gr/Ir}. An Ir single crystal cut in the (111) surface purchased from Surface Preparation Laboratory was cleaned in the same UHV chamber as for gr/Re, by cycles of Ar$^+$ ion bombardment at $1~\mathrm{keV}$ and subsequent annealing at $1,200~\degree\mathrm{C}$.
The gr/Ir was prepared by exposing to $10^{-8}~\mathrm{mbar}$ of $\mathrm{C_2H_4}$ at $1,000~\degree\mathrm{C}$ for 15~minutes.

\noindent \textbf{STM measurements}. For multilayer gr/SiC, STM measurements were performed at $4~\mathrm{K}$ in a home-made He-cooled STM, using a commercial Pt/Ir tip bought from Bruker. For gr/Re and gr/Ir, STM measurements were performed at room temperature under UHV, using a commercial Omicron UHV-STM~1, with a W chemically etched tip. Before analysing STM images, thermal drift and miscalibrations have been corrected.

\noindent \textbf{DFT calculations}. DFT calculations were performed using the VASP code, with the projector augmented wave (PAW) approach\cite{kresse_ab_1993,kresse_ultrasoft_1999}{}. The exchange correlation interaction is treated within the general gradient approximation parameterized by Perdew, Burke and Ernzerhof (PBE)\cite{perdrew_general_1996}{}. The Methfessel Paxton method is used to calculate the total energy with a smearing of $0.2$. The cut-off energy is of $400$~eV. The supercell consists in four Re layers and one C layer with an empty space of $9~\text{\AA}$ to avoid spurious interactions. Re atoms are kept fixed in the bottom second Re layer, all other atoms are allowed to relax. Due to the size of the supercell, calculations are performed using the K point only. After convergence, residual forces are lower than $0.03$~eV/$\text{\AA}$.

\section*{Acknowledgements}

We thank Jean-Yves Veuillen and Karim Ferhat for fruitful discussions. T.L.Q. was supported by a CIBLE fellowship from R$\text{\'e}$gion Rh$\text{\^o}$ne-Alpes.

\section*{Author contributions statement}

C.C. and J.C. conceived the experiments,  A.A., T.L.Q., V.G. and P.D. conducted the experiments, L.M. performed the simulations, A.A. analysed the results. All authors reviewed the manuscript. 

\section*{Additional information}

\textbf{Competing financial interests}: The authors declare no competing financial interests. 

\end{document}